\documentclass[12pt]{iopart}
\usepackage{bm}
\usepackage{epsfig}
\usepackage{citesort}
\eqnobysec

\begin{document}

\begin{flushright}
{\large  MPP-2006-82}
\end{flushright}

\title{Neutrino masses and cosmic radiation density:
Combined analysis}

\author{Steen Hannestad$^{1,2}$ and Georg G.~Raffelt$^2$}

\address{$^1$ Department of Physics and Astronomy,
University of Aarhus, DK-8000 Aarhus C, Denmark}

\address{$^2$Max-Planck-Institut f\"ur Physik
(Werner-Heisenberg-Institut),
F\"ohringer Ring 6, D-80805 M\"unchen, Germany}

\ead{\mailto{sth@phys.au.dk}, \mailto{raffelt@mppmu.mpg.de}}

\begin{abstract}
We determine the range of neutrino masses and cosmic radiation
content allowed by the most recent CMB and large-scale structure
data. In contrast to other recent works, we vary these parameters
simultaneously and provide likelihood contours in the
two-dimensional parameter space of $N_{\rm eff}$, the usual
effective number of neutrino species measuring the radiation
density, and $\sum m_\nu$. The allowed range of $\sum m_\nu$ and
$N_{\rm eff}$ has shrunk significantly compared to previous
studies. The previous degeneracy between these parameters has
disappeared, largely thanks to the baryon acoustic oscillation
data. The likelihood contours differ significantly if $\sum m_\nu$
resides in a single species instead of the standard case of being
equally distributed among all flavors. For $\sum m_\nu=0$ we find
$2.7<N_{\rm eff}<4.6$ at 95\% CL while $\sum m_\nu<0.62$~eV at
95\% CL for the standard radiation content.
\end{abstract}

\maketitle

\section{Introduction}

The recent release of the 3-year WMAP data has stimulated several
renewed analyses of cosmological neutrino mass limits. Neutrinos are
known to have mass from oscillation experiments so that the unknown
overall mass scale is unavoidable as a fit parameter of the standard
cosmological model. The resulting mass limits range from $\sum m_\nu<
2.0$~eV (95\% CL) using the WMAP-3 data alone~\cite{fukugita} to $\sum
m_\nu< 0.17$--$0.4$~eV (95\% CL) when data from the Lyman-$\alpha$
forest is included~\cite{seljak2006,Goobar:2006xz}.

Translating cosmological limits on the hot dark matter fraction into
neutrino mass limits depends on the cosmic neutrino density that is
fixed by standard physics and thus not an ordinary cosmic fit
parameter. On the other hand, the most direct evidence for the
presence of the cosmic neutrino sea derives from big-bang
nucleosynthesis and from cosmological parameter fitting so that it
is a natural consistency test to study if the cosmic radiation
density implied by the cosmological precision parameters reproduces
the standard radiation content.

However, since neutrinos are known to have mass, one can not
simply assume that $\sum m_\nu=0$ when extracting an allowed range
for $N_{\rm eff}$, the effective number of neutrino species that
is the usual measure of the radiation content. Nevertheless this
has been the standard procedure in most of the recent parameter
analyses based on WMAP-3 \cite{Spergel:2006hy,seljak2006}. The
caveat is especially relevant because we found a degeneracy
between $\sum m_\nu=0$ and $N_{\rm eff}$, based on the
cosmological data available in 2003
\cite{Hannestad:2003xv,Hannestad:2003ye,Crotty:2004gm,Dodelson:2005tp}.
One result of our present $\sum m_\nu$-$N_{\rm eff}$-analysis will
be that this degeneracy is no longer present in the much smaller
allowed range based on the 2006 data (Fig.~\ref{fig:old}). We will
thus conclude that the current cosmological data provide
essentially independent limits on $\sum m_\nu$ and $N_{\rm eff}$.

\begin{table}[b]
 \begin{center}
 \begin{tabular}{llll}
 \hline\hline
 Case & Model & $\sum m_\nu$ (95\%~CL)& $N_{\rm eff}$ (95\%~CL)\cr
 \hline 1 & $N_{\rm m} = N_{\rm eff}$ & $<$ 0.62 eV & $2.7 <  N_{\rm eff} < 4.6$ \cr
 \hline
 2 & $N_{\rm m} = 3$ & $<$ 0.57 eV & $3.0 \leq  N_{\rm eff} < 4.6$ \cr
 \hline
 3 &$N_{\rm m} = 1$ & $<$ 0.41 eV & $2.7 <  N_{\rm eff} < 4.6$ \cr
 \hline\hline
 \end{tabular}
 \end{center}
\caption{Cases of neutrino mass distribution among the $N_{\rm eff}$
species. For the allowed $\sum m_\nu$ range we have marginalized over
$N_{\rm eff}$ and vice versa.}\label{table:cases}
\end{table}

Another important issue is what one actually means with $N_{\rm
eff}$, as there are several different plausible cases that should
be considered. One possibility is that the cosmic number density
of the standard neutrinos is different from what is usually
assumed, i.e.\ $\sum m_\nu$ is equally distributed among all
species that comprise~$N_{\rm eff}$ (our Case~1, see
Table~\ref{table:cases}).  Our second case is that three standard
massive neutrinos have equal masses, i.e.\ the number of equally
massive species is $N_{\rm m}=3$ so that $N_{\rm eff}-N_{\rm m}$
signifies additional radiation in some completely new form
unrelated to ordinary neutrinos.  Finally, we consider $N_{\rm
m}=1$ that could represent a situation where the standard
neutrinos are nearly massless, i.e.\ they have hierarchical masses
with a largest mass eigenvalue given by the atmospheric scale of
about 50~meV, while there is an additional massive species,
perhaps a sterile neutrino.  This case is largely motivated to
demonstrate that the current cosmological data are sensitive to
the $\sum m_\nu$ distribution among the flavors.

We begin in Sec.~\ref{sec:data} with a brief description of the
cosmological data used in our study.  In Sec.~\ref{sec:constraints} we
derive the range of $\sum m_\nu$ and $N_{\rm eff}$ allowed by these
data and conclude in Sec.~\ref{sec:summary} with a summary of our
findings.

\section{Cosmological data and likelihood analysis}
\label{sec:data}

In order to study bounds on $\sum m_\nu$ and $N_{\rm eff}$ we use the
same data as in Ref.~\cite{Goobar:2006xz}. We use distant type~Ia
supernovae measured by the SuperNova Legacy Survey
(SNLS)~\cite{astier06} and large-scale structure data from the
2dF~\cite{2dFGRS} and SDSS~\cite{Tegmark:2003uf,Tegmark:2003ud}
surveys. From the SDSS we also include the recent measurement of the
baryon acoustic oscillation feature in the 2-point correlation
function~\cite{Eisenstein2005}.  Finally, we include the precision
measurements of the cosmic microwave background anisotropy from the
WMAP experiment~\cite{Spergel:2006hy,Hinshaw:2006ia,Page:2006hz}, as
well as the smaller-scale measurement by the BOOMERANG
experiment~\cite{Jones:2005yb,Piacentini:2005yq,Montroy:2005yx}.

\begin{table}[b]
\begin{center}
\begin{tabular}{lcl}
\hline \hline parameter & prior\cr
\hline $\Omega=\Omega_m+\Omega_{\rm DE} + \Omega_\nu$&1&Fixed\cr
$\Omega_m$ & 0 -- 1 & Top hat \cr
 $h$ & 0.5 -- 1.0 & Top hat
\cr $\Omega_b h^2$ & 0.014 -- 0.040&Top hat\cr  $w_{\rm DE}$ &
$-2.5$ -- $-0.5$ & Top hat \cr $n_s$ & 0.6 -- 1.4& Top hat\cr
$\alpha_s$ & $-0.5$ -- 0.5 & Top Hat \cr $\tau$ & 0 -- 1 &Top hat\cr
$Q$ &
--- &Free\cr $b$ & --- &Free\cr
\hline \hline
\end{tabular}
\end{center}
\caption{Priors on the parameters used in our likelihood
analysis.} \label{table:priors}
\end{table}

We do not include data from the Lyman-$\alpha$ forest in our
analysis. These data were used previously and very strong separate
bounds on $\sum m_\nu$ and $N_{\rm eff}$ were
obtained~\cite{seljak2006}.  However, the strength of these bounds is
mainly related to the fact that the Lyman-$\alpha$ analysis used in
Ref.~\cite{seljak2006} leads to a much higher normalisation of the
small-scale power spectrum than the WMAP data.  Other analyses of the
same SDSS Lyman-$\alpha$ data find a lower normalisation, in better
agreement with the WMAP
result~\cite{Viel:2005eg,Viel:2005ha,viel2006}.  In this case the
Lyman-$\alpha$ data add little to the strength of the neutrino mass
bound~\cite{Goobar:2006xz}. The discrepancy between different analyses
of the same data probably points to unresolved systematic issues so
that we prefer to exclude the Lyman-$\alpha$ data entirely.

We then perform a likelihood analysis based on a flat, dark-energy
dominated model characterised by the matter density $\Omega_m$, the
baryon density $\Omega_b$, the dark energy equation of state $w$, the
Hubble parameter $H_0$, the spectral index of the primordial power
spectrum $n_s$, the running of the primordial spectral index
$\alpha_s$, and the optical depth to reionization $\tau$.  Finally,
the normalization of the CMB data $Q$ and the bias parameter $b$ are
used as free parameters. The dark-energy density is given by the
flatness condition $\Omega_{\rm DE} = 1 - \Omega_m -
\Omega_\nu$. Including the neutrino mass $\sum m_\nu$, parameterised
in terms of the contribution to the present energy density
$\Omega_\nu h^2 = \sum m_\nu/92.8~{\rm eV}$, and the effective number
of neutrino species $N_{\rm eff}$, our benchmark model has 11 free
parameters.

Our priors on these parameters are shown in
Table~\ref{table:priors}. The treatment of data is exactly the same as
in Ref.~\cite{Goobar:2006xz}.  When calculating constraints, the
likelihood function is found by minimizing $\chi^2$ over all
parameters not appearing in the fit, i.e.~over all parameters other
than $\sum m_\nu$ and $N_{\rm eff}$.

\begin{figure}[b]
\vspace*{-0.0cm}
\begin{center}
\epsfxsize=7.5truecm\epsfbox{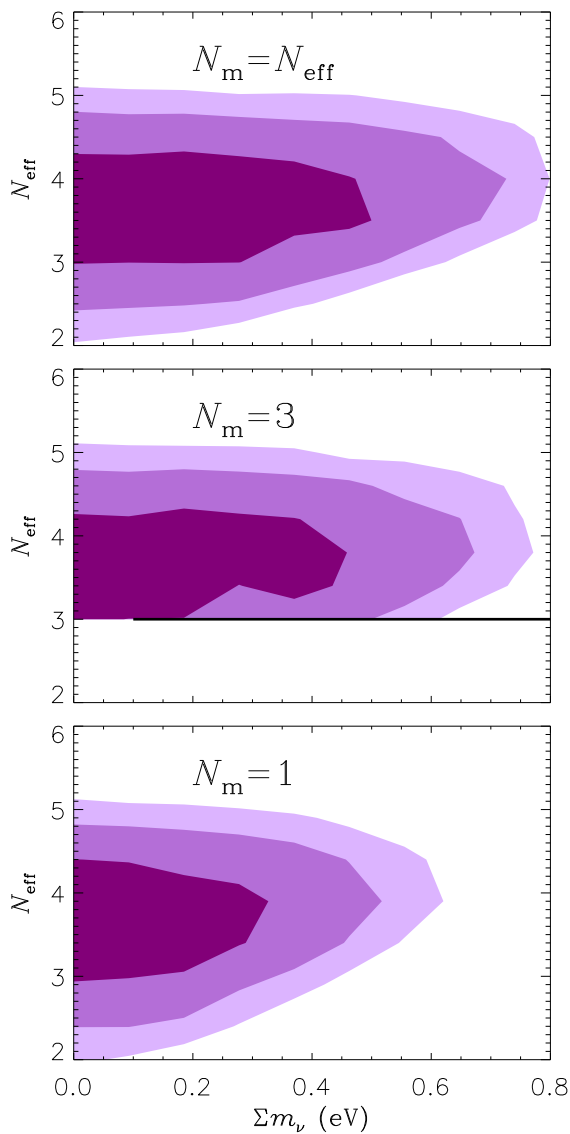} \vspace{0.5truecm}
\end{center}
\caption{The 68\%, 95\%, and 99\% confidence level contours for
our three cases.} \label{fig:likelihood}
\end{figure}

\section{Bounds on neutrino properties}        \label{sec:constraints}

\begin{figure}[t]
\vspace*{-0.0cm}
\begin{center}
\epsfxsize=7.5truecm\epsfbox{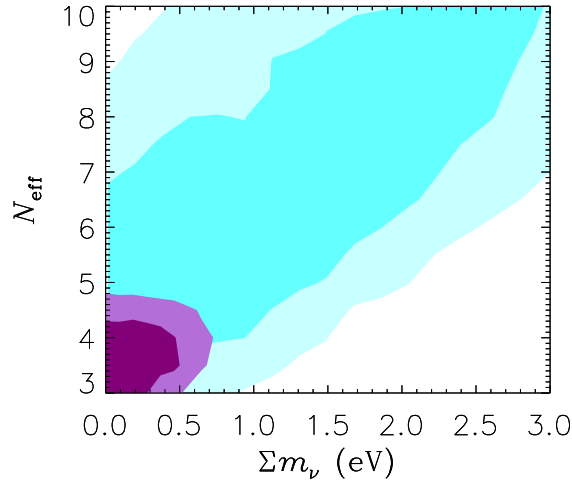} \vspace{0.5truecm}
\end{center}
\caption{The 68\% and 95\% confidence level contours for Case~1
overlayed on the same contours using the data available in
2003~\cite{Hannestad:2003ye}.}\label{fig:old}
\end{figure}

Following these procedures we find the 68\%, 95\%, and 99\%
likelihood contours for $\sum m_\nu$ and $N_{\rm eff}$ shown in
Fig.~\ref{fig:likelihood} for the three cases discussed in the
introduction and shown in Table~\ref{table:cases}. The top panel of
Fig.~\ref{fig:likelihood} corresponds to a nonstandard number
density of the standard neutrinos, assuming a standard velocity
dispersion as in all other cases as well. In Fig.~\ref{fig:old} we
overlay these contours with the analogous ones that we found on the
basis of the data available in 2003~\cite{Hannestad:2003ye}.  The
allowed range of both parameters has shrunk dramatically as
expected. Moreover, the pronounced degeneracy between $\sum m_\nu$
and $N_{\rm eff}$ that was present at that time has now completely
disappeared, largely thanks to the baryon acoustic oscillation (BAO)
measurements. We conclude that at the level of precision that has
now been reached, the cosmological data constrain $\sum m_\nu$ and
$N_{\rm eff}$ almost independently of each other.

Perhaps the physically best motivated case is No.~2 where we have the
ordinary neutrinos with mass ($N_{\rm m}=3$) and additional radiation
in some new form. In this case we have a hard lower limit $N_{\rm
eff}\geq N_{\rm m}=3$. Otherwise the contours of the middle panel of
Fig.~\ref{fig:likelihood} are very similar to Case~1
(top panel).

The largest modification appears in Case~3 where we assume that
all hot-dark matter mass resides in a single neutrino species. The
mass limits are significantly more restrictive in this case. The
reason is that for a single massive species the total neutrino
energy density is larger in the semi-relativistic regime than if
the mass is shared between all flavours \cite{Lesgourgues:2004ps}.
Since the mass bound is such that neutrinos become nonrelativistic
very close to the epoch of matter-radiation equality, this
equality occurs later in the model with only one massive neutrino.
As a consequence, small-scale structure is more suppressed, but
the effect can be offset by a slight increase in the matter
density.  We indeed observe that the best-fit value of $\Omega_m$
is higher for $N_{\rm m}=1$.  However, both the SN~Ia and BAO data
prefer a low value of $\Omega_m$ and consequently the model with
$N_{\rm m}=1$ becomes a poor fit to this data at $\sum m_\nu$
around 0.4--0.5~eV.  Table~\ref{table:deltachi} shows exactly this
effect. Here, $\Delta \chi^2$ for $\sum m_\nu = 0.45$~eV has been
broken down into individual contributions from the different data
sets. As expected, the main effect comes from SN~Ia and BAO data.

\begin{table}[ht]
\begin{center}
\begin{tabular}{lcc}
\hline \hline Data set & $N_m = N_{\rm eff}$ & $N_m = 1$ \cr
\hline CMB & $-1.2$ & $-0.9$ \cr LSS & 0.5 & 0.5 \cr SN~Ia & 1.0 & 2.3
\cr BAO & 1.3 & 2.9 \cr \hline \hline
\end{tabular}
\end{center}
\caption{$\Delta \chi^2$ for $\Omega_\nu h^2 = 0.005$ compared
with the best fit model, broken down into individual
contributions.} \label{table:deltachi}
\end{table}

We finally note that for the case with one sterile massive state
and three active, almost massless neutrinos (the LSND 3+1 case)
the mass bound is 0.45 eV at 95\% CL (0.93 eV at 99.99\% C.L.),
somewhat lower than the 0.62 eV bound in the standard case. That
the bound on the 3+1 model is stronger than for the standard case
is contrary to what was found in previous studies (see
\cite{Hannestad:2003xv,Hannestad:2003ye,Crotty:2004gm}. The reason
is the low value of $\Omega_m$ preferred by the BAO and SNI-a
data.

\section{Discussion}                               \label{sec:summary}

We have derived likelihood contours in the two-dimensional
parameter space spanned by $\sum m_\nu$ and $N_{\rm eff}$, based
on the latest cosmological precision data, however excluding
Lyman-$\alpha$. We consider two physically motivated cases for
$N_{\rm eff}$ where either the effective number of massive
neutrinos differs from the standard scenario, or where there is a
new form of radiation besides $N_{\rm m}=3$ standard massive
neutrinos. The results for these cases differ very little, except
that in Case~2 there is hard lower limit $N_{\rm eff}\geq N_{\rm
m}=3$.

For the sake of principle we have also considered a third case where
all the neutrino mass resides in a single species. Here, the mass
limit is more restrictive, reflecting that near the limiting mass
of around 0.5~eV neutrinos become nonrelativistic very close to the
epoch of matter-radiation equality.

For all cases we provide in Table~\ref{table:cases} limits on $\sum
m_\nu$ after marginalizing over $N_{\rm eff}$ and limits on $N_{\rm
eff}$ after marginalizing over $\sum m_\nu$. We stress that the
neutrino mass scale can not be avoided as a standard cosmic fit
parameter so that one should not derive limits on $N_{\rm eff}$ while
enforcing the neutrino masses to vanish. In practice, because $\sum
m_\nu$ and $N_{\rm eff}$ are no longer degenerate, the allowed range
for $N_{\rm eff}$ is not very different if one assumes $\sum m_\nu=0$.

The limits on $N_{\rm eff}$ differ little between our cases.
Independently of the exact distribution of masses among the
neutrino species we find $2.7 < N_{\rm eff} < 4.6$ (95\%~CL),
except in Case~2 where the lower limit is by definition $3\leq
N_{\rm eff}$. The bound is significantly stronger than the $2.5 <
N_{\rm eff} < 5.6$ found with WMAP-1 data and without inclusion of
the BAO data \cite{Hannestad:2005jj}. The standard case of three
massive neutrinos without modified number density and without
additional radiation is well within the 95\% CL range of $N_{\rm
eff}$, although it is just slightly outside the 68\% CL range in
Cases~1 and 2. Either way, the cosmological model with a
nonstandard density of massive neutrinos or radiation is not
significantly favored over the standard case.

\section*{Acknowledgments}

This work was supported, in part, by the Deutsche
Forschungsgemeinschaft under grant No.~SFB 375 and by the European
Union under the ILIAS project, contract No.~RII3-CT-2004-506222.
S.H.\ acknowledges support from the Alexander von Humboldt
Foundation through a Friedrich Wilhelm Bessel Award. Use of the
CMBFAST code is acknowledged \cite{Seljak:1996is}.

\section*{References}

\end{document}